\documentclass[aps,prl,showpacs,twocolumn,superscriptaddress]{revtex4}

\usepackage{amsmath}
\usepackage{graphicx}
\usepackage{color}
\usepackage{amssymb}

\newcommand{\VS}{V_N}
\newcommand{\IS}{I_N}

\newcommand{\ocite}{\cite}

\begin{document}
\title{Measuring non-Gaussian fluctuations through incoherent Cooper pair current}

\author{Tero T. Heikkil\"a}
\email{Tero.T.Heikkila@hut.fi}

\affiliation{Low Temperature Laboratory, P.O. Box 2200, FIN-02015
HUT, Finland}

\affiliation{Department of Physics and Astronomy, University of
Basel, Klingelbergstr. 82, CH-4056 Basel, Switzerland}

\author{Pauli Virtanen}

\affiliation{Low Temperature Laboratory, P.O. Box 2200, FIN-02015
HUT, Finland}

\author{G\"oran Johansson}

\affiliation{Institut f\"ur Theoretische Festk\"orperphysik,
Universit\"at Karlsruhe, D-76128 Karlsruhe, Germany}

\affiliation{Applied Quantum Physics, MC2, Chalmers University of
Technology, S-412 96 G\"oteborg, Sweden}

\author{Frank K. Wilhelm}

\affiliation{Physics Department and CeNS,
Ludwig-Maximilians-Universit\"at, Theresienstr. 37, D-80333
M\"unchen, Germany}

\date{\today}


\begin{abstract}
We study a Josephson junction (JJ) in the regime of incoherent
Cooper pair tunneling, capacitively coupled to a nonequilibrium
noise source. The current-voltage (I-V) characteristics of the JJ
are sensitive to the excess voltage fluctuations in the source,
and can thus be used for wide-band noise detection. Under weak
driving, the odd part of the I-V can be related to the second
cumulant of noise, whereas the even part is due to the third
cumulant. After calibration, one can measure the Fano factors for
the noise source, and get information about the frequency
dependence of the noise.
\end{abstract}

\maketitle

The current in electric circuits fluctuates in time, even when
driven with a constant voltage. At equilibrium or in large
conductors, this current noise can be quantified using the
fluctuation-dissipation theorem (FDT), which relates the magnitude
of the fluctuations to the temperature $T$ and the impedance of
the circuit. Moreover, in large wires the current statistics is
described by a Gaussian probability density which has only two
nonzero cumulants, the average current and noise power. This
situation changes for small, mesoscopic-scale resistors exhibiting
shot noise \cite{blanterbuettiker,nazarovnoise}: The noise power
at low frequencies is proportional to the average current.
Further, the statistics of the transmitted charge is no longer
Gaussian: higher cumulants are finite, and the probability density
is "skew", i.e., odd cumulants do not vanish.

For small samples, the frequency scale for the shot noise is given
by the voltage, $eV/\hbar$. Shot noise has been measured at low
frequencies in many types of mesoscopic structures (see the
references in \cite{blanterbuettiker,nazarovnoise}). However,
there are only a few direct measurements of shot noise at high
frequencies $\omega \sim eV/\hbar$ \cite{hfnoise}, and only one of
the higher (than second) cumulants \cite{reulet03} (at $\omega \ll
eV/\hbar$). One of the main reasons for the shortage of such
measurements is the difficulty to couple the fluctuations to the
detector at high frequencies, or to devise wide-band detection, as
required for the third and higher cumulants
\cite{reulet03,heikkilaup04}.

In this Letter, we analyze an on-chip detector of voltage
fluctuations, based on capacitively coupling a noise source to a
small JJ in Coulomb blockade \cite{delahaye03}. There, the current
can flow only if the environmental fluctuations provide the
necessary energy to cross the blockade. In this way, the current
through the small JJ provides detailed information of the voltage
fluctuations in the source over a wide bandwidth. This information
includes effects of a non-Gaussian (``skew'') environment on a
quantum system. For the measurement of the second cumulant, its
characteristics compare well with the other suggested on-chip
detectors
\cite{deblock03,aguado00,schoelkopf03,sonin04,tobiskapekola},
based on various mesoscopic devices and techniques. The detectors
proposed in \cite{sonin04,tobiskapekola} detect the non-Gaussian
character of the noise, but mapping the output back to the
different cumulants has not been carried out. The on-chip scheme
presented here is the first to directly measure the third cumulant
of fluctuations. In the Gaussian regime, our analysis of the noise
detection resembles that of Ref.~\ocite{aguado00}, but probes the
noise at low measurement voltages.

\begin{figure}[h]
\centering
\includegraphics[width=\columnwidth]{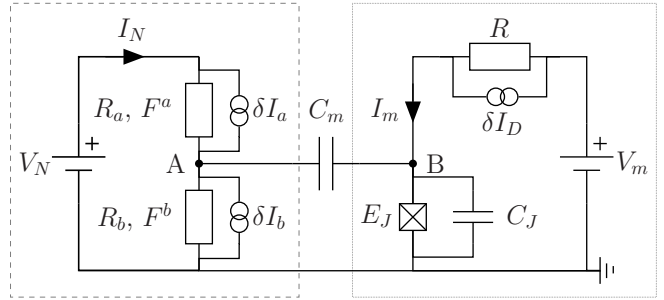}
\caption{Noise measurement circuit. The voltage noise at point A,
driven with $\VS$, induces voltage noise at point B through the
capacitor $C_m$ and in this way affects JJ. JJ is in a highly
resistive environment, $R
> R_Q=h/(4e^2)$. $F_n^{a,b}$ are the Fano
factors for the $n$'th cumulants in the noise source. The noise is
read by investigating $I_m(V_m)$, as explained in the text.}
\label{fig:circuit}
\end{figure}

% What the system looks like
We investigate the system depicted in Fig.~\ref{fig:circuit}. The
part indicated by the dashed lines represents the noise source,
and the other part is the detector. The excess voltage noise at
point $A$, induced by driving the source ($\VS \neq 0$), gives
rise to voltage noise at point $B$, and the latter can be read by
examining the current $I_m(V_m)$. There is also another source of
noise: the equilibrium fluctuations in the whole circuit,
described by FDT.
As shown below, the two types of fluctuations %are uncorrelated and
can be treated separately.

In the regime of incoherent Cooper-pair tunneling, the I-V
characteristics of the detector can be described by a perturbation
theory in the Josephson coupling energy $E_J$. This yields for the
current through JJ \cite{ingoldnazarov,ingold99}
\begin{equation}
I_m(V_m)=\frac{\pi e E_J^2}{\hbar}
\left[P_{\phi(t)}(2eV_m)-P_{-\phi(t)}(-2eV_m)\right].
\end{equation}
Here $P_{\phi(t)}(E)\equiv \int dt e^{iEt/\hbar} \left\langle e^{i
\phi(t)} e^{-i \phi(0)} \right\rangle/(2\pi \hbar)$ describes the
Cooper-pair tunneling through the JJ due to the fluctuations
$\phi(t)$ of the phase difference across the junction. This result
makes no assumptions about the form of the phase fluctuations -
only that they do not modify $E_J$ itself. As these fluctuations
are connected to the fluctuations $\delta V_B(t)$ of the voltage
over the JJ via the Josephson relation, $\phi(t)=\phi(0)+2e
\int_0^t dt' \delta V_B(t')/\hbar$, $I_m(V_m)$ provides
information on these fluctuations. Note that, although the
coupling capacitance acts as a high-pass filter for voltage noise,
the conversion to phase noise allows to transmit noise down to low
frequencies. This way the device can be operated
non-hysteretically \cite{sonin04} and in Coulomb blockade.

One can identify $P_\phi(E)$ as the Fourier transform of the
moment-generating function $\kappa_\phi(\chi)=\langle e^{\chi
\phi(t)} e^{-\chi \phi(0)}\rangle$ of $\phi(t)-\phi(0)$, evaluated
at $\chi=i$. This can be expanded in the cumulants $C_n^\phi(t)$,
$\kappa_\phi(\chi)=\exp[\chi^2/(2!) C_2^\phi(t)+\chi^3/(3!)
C_3^\phi(t) + \dotsb]$. These cumulants are defined such that
$\phi(t)$ is ordered before $\phi(0)$ in the expectation value.
The expansion defines a function
$J_\phi(t)=\ln(\kappa_\phi(i))=J_2(t)+J_3(t)+\dotsb$, where for
stationary fluctuations $J_2(t)=\langle
[\phi(t)-\phi(0)]\phi(0)\rangle$ and $J_3(t)=\frac{i}{2}\langle
\phi(t)[\phi(t)-\phi(0)]\phi(0) \rangle$, etc. In the Gaussian
limit, $J_\phi(t)=J_{-\phi}(t)=J_2(t)$ coincides with that applied
in Ref.~\ocite{ingoldnazarov}.

The odd cumulants of the fluctuations in the source break the
symmetry between the positive and the negative $V_m$. To separate
the non-Gaussian effects, we consider the even and odd parts of
the I-V, $I_{S/A}(V_m) \equiv (I_m(V_m) \pm I_m(-V_m))/2$. The odd
part $I_A(V_m)$ describes the general behavior mostly due to the
even cumulants (c.f., Eq.~\eqref{eq:antisymmcur} below), and the
even part $I_S(V_m)$ (Eq.~\eqref{eq:symcur}) responds to the odd
cumulants, vanishing if $C_{2n+1}=0$. We show below that tuning
the voltage $V_m$ and measuring $I_{A/S}(V_m)$ gives access to the
frequency dependence of the lowest cumulants at large bandwidths.

We have to take the additional Gaussian fluctuations $J_\phi^D(t)$
from the total impedance of the setup into account, as they will
inevitably influence the measurement. Thus, we split the
fluctuations as parts $J_\phi(t)=J_\phi^D(t)+J_\phi^S(t)$ with the
excess fluctuations due to the driven source denoted by
$J_\phi^S(t)$. The latter includes the non-Gaussian effects for
which $J_{-\phi}^S \neq J_\phi^S$. We also define $P^{D/S}_\phi(E)
\equiv \int dt e^{iEt/\hbar} \langle e^{J_\phi^{D/S}(t)}
\rangle/(2\pi \hbar)$. Using detailed balance
\cite{ingoldnazarov}, we obtain for the current $I_D(V_m)\equiv
I_m(V_m;J_\phi^S=0)=\pi e E_J^2 [P_\phi^D(2eV_m)(1-\exp(-\beta 2
eV_m))]/\hbar$. In the presence of the excess noise, the current
is given by
\begin{equation}
I_m = \left[\frac{I_D\left(\frac{E}{2e}\right)}{1-e^{-\beta E}} *
\left(P_\phi^S(E)-e^{\beta (E-2eV_m)} P_{-\phi}^S(-E)
\right)\right],
\end{equation}
evaluated at $E=2eV_m$. Here $\beta^{-1}=k_B T$ and $*$ denotes
the convolution over the energy. Thus, the detector
characteristics, described by $P^D_\phi(E)$, do not need to be
known exactly, but they can be calibrated by measuring the current
$I_D(V_m)$ in the absence of the additional noise.

For connecting the fluctuations at the JJ and the source, we
relate the mutually uncorrelated intrinsic fluctuations $\delta
I_a$, $\delta I_b$ and $\delta I_m$ through the resistors $R_a$,
$R_b$ and $R$ to the voltage fluctuations $\Delta V$ at point $B$,
$\Delta V(\omega)=R[-\delta I_m(\omega) - i \omega C_m R_S (\delta
I_b-\delta I_a-\delta I_m)]/G(\omega)$ through circuit analysis.
Here $G(\omega)=1-i\omega (R C + R_S C_m)-\omega^2 R C_J R_S C_m$,
$C=C_m+C_J$ and $R_S=(R_a^{-1}+R_b^{-1})^{-1}$. The equilibrium
fluctuations in the source are present even for $\VS=0$, and they
can be included in the calibration through the FDT. The excess
fluctuations due to driving produce excess phase fluctuations,
characterized by the $n$'th order correlators $\langle
\phi(\omega_1)\dotsm \phi(\omega_n)\rangle \equiv 2\pi
\delta(\omega_1+\dots+\omega_n)S_{n\phi}(\Vec{\omega})$
\cite{extracorrnote},
\begin{equation}
S_{n\phi}(\Vec{\omega}) =\lambda^n \frac{\tau^n}{e^n}
\frac{S_{nI}^b(\Vec{\omega}) + (-1)^n
S_{nI}^a(\Vec{\omega})}{G_n(\Vec{\omega})}, \label{eq:phicum}
\end{equation}
where $\Vec{\omega}=(\omega_1,\dots, \omega_n)$, $\lambda=\pi R_S
C_m/(R_Q C)$, $R_Q=h/(4e^2)$, $\tau=RC$,
$G_n(\Vec{\omega})=G(\omega_1) \dotsm G(\omega_n)$, and $\langle
\delta I(\omega_1) \dotsm \delta I(\omega_n) \rangle=2\pi
\delta(\omega_1 + \dots \omega_n)S_{nI}(\vec{\omega})$. Thus, we
find that the excess phase noise and its cumulants are governed by
powers of $\lambda$, the current $\IS \tau/e$ and the bandwidth of
the current correlators, times $\tau$. In the absence of the
source ($C_m \rightarrow 0$), the detector phase fluctuations
$J_{RC}(t)$ have been calculated in \cite{grabert98} (see also
Fig.~11 of \cite{ingoldnazarov}). The equilibrium fluctuations in
the source slightly modify this behavior \cite{virtanenup}. If
$R_S C_m \ll R C_J$, the resulting $J^D(t)$ is close to
$J_{RC}(t)$, with the capacitance given by $C_m+C_J$ and
resistance by $R$, hence the JJ is in the insulating state for $R
> R_Q$. Below, we concentrate on this insulating parameter regime.

Now assume the source is driven weakly, such that $J^S(t) \ll 1$.
In this case, one may expand the exponential $e^{J^S(t)} \approx 1
+ J^S(t)$ \cite{expansionnote}. Then,
\begin{equation}
P_\phi^S(E)=\delta(E)+\frac{S_{2\phi}(E/\hbar)}{2\pi
\hbar}-\tilde{S}_{2\phi}(0) \delta(E)+\frac{K_\phi(E/\hbar)}{4\pi
\hbar} +\dotsb. \label{eq:pefromexp}
\end{equation}
Here $S_{2\phi}(\omega)$ is the driven phase noise spectrum
induced by the source, $\tilde{S}_{2\phi}(t)$ its inverse Fourier
transform, and
\begin{equation}
K_\phi(\omega) = \frac{1}{\pi} \int_{-\infty}^\infty d \omega'
{\rm Im} S_{3\phi}(\omega,\omega',-\omega-\omega')
\end{equation}
describes the third cumulant of phase fluctuations. In the
derivation of this form for $K_{\phi}$, we used the hermiticity of
$\phi(t)$ and the stationarity of the fluctuations. In what
follows, we cut the expansion in the third cumulant
\cite{expansionnote}.

\begin{figure}[h]
\centering \includegraphics[width=\columnwidth]{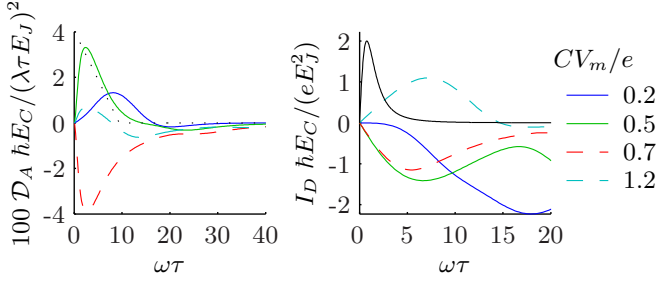}
\caption{(Color online) Left: characteristic function ${\cal
D}_A(\omega;V_m)$ of the excess noise measurement, (c.f.,
Eq.~(\ref{eq:antisymmcur})), probing the spectrum $S_{2I}(\omega)$
of current fluctuations. Dotted line shows the frequency
dependence in $S_{2I}(\omega)$ (arbitrary units), for $T=0$ and
$V=e/C$. Right: function $[I_D(V_m+\hbar
\omega/(2e))-I_D(V_m-\hbar \omega/(2e))]/2$, characterizing the
measurement of the third cumulant. The frequency dependence in
$K_\phi(\omega)$ (arbitrary units) is shown in black. In the limit
$e\VS \gg \hbar/\tau$, the width of $K_\phi(\omega)$ is given by
$1/\tau$. Here, $E_C \equiv 2e^2/C$, $R=6\,R_Q$,
$R_a=R_b=0.1\,R_Q$, and $C_m=10C_J$. Changes in $C_m$ do not
essentially affect the figure.} \label{fig:detectorband}
\end{figure}

Let us first concentrate on the antisymmetric part of the detector
current. In the first order in $J^S(t)$, it only probes the even
cumulants. In this case, it can be expressed as $I_A(V_m) \equiv
I_D(V_m)+\delta I_A(V_m)$, where for symmetric
$S_{2I}(\omega)=S_{2I}(-\omega)$ \cite{neglectedpart}, using
Eq.~\eqref{eq:phicum},
\begin{equation}
\delta I_A(V_m)=\int_{-\infty}^\infty \frac{d\omega}{2\pi} {\cal
D}_A(\omega;V_m) (S_{2I}^b(\omega)+S_{2I}^a(\omega)).
\label{eq:antisymmcur}
\end{equation}
Here ${\cal D}_A(\omega;V_m)\equiv\lambda^2 \tau^2[I_D(V_m+\hbar
\omega/(2e))+I_D(V_m-\hbar \omega/(2e))-2I_D(V_m)]/(2 e^2
G_2(\omega))$ characterizes the frequency band for the detection
of $S_{2I}(\omega)$. This band can be tuned by tuning the bias
$V_m$ (see Fig.~\ref{fig:detectorband}).

The even part of the detector current can then be related to the
odd cumulants of phase fluctuations. For a weakly driven source,
assuming $K_\phi(-\omega)=-K_\phi(\omega)$ (see below), it is
given by \cite{neglectedpart}
\begin{equation}
I_S(V_m)= \int_{-\infty}^\infty \frac{d\omega}{4\pi}
I_D\left(V_m-\frac{\hbar \omega}{2e}\right)
K_\phi\left(\omega\right). \label{eq:symcur}
\end{equation}
In this way, $I_S(V_m)$ probes the frequency dependence of the
third cumulant of source fluctuations, (c.f.,
Fig.~\ref{fig:detectorband}).

\begin{figure}\centering
\includegraphics[width=0.93\columnwidth]{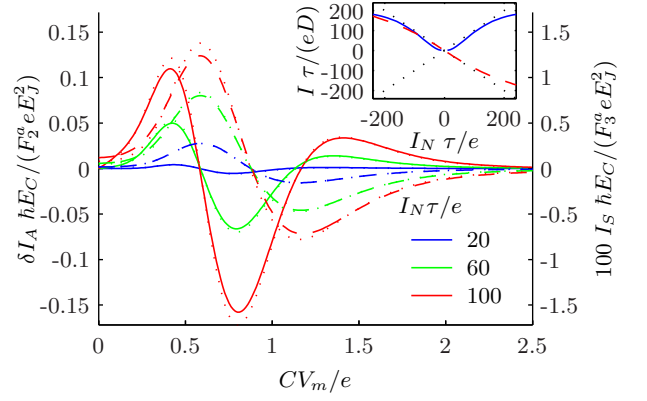}
\caption{(Color online) Antisymmetric $\delta{}I_A(V_m)$ (solid,
left axis) and symmetric part $I_S(V_m)$ (dashed, right axis) of
the detector current variation. Dotted lines are expansion results
(Eqs.~\eqref{eq:antisymmcur}, \eqref{eq:symcur}). Parameters are
as in Fig.~\ref{fig:detectorband}. The source is assumed to
consist of a mesoscopic ($F_n^a \neq 0$) and a macroscopic
resistor ($F_n^b=0$) for which $S_{nI}(\omega=0)=e^{n-1} F_n I_N$.
Inset: $\delta I_A(V_m=e/2C)$ (solid) and $I_S(V_m=e/2C)$ (dashed)
as functions of $I_N$. Scaling parameters are $D=F_2^a
D_A\equiv{}eF_2^a\int\mathrm{d}\omega{}{\cal
D}_A(\omega;V_m)/(2\pi{})$ for the antisymmetric part, and
$D=F_3^a
D_S\equiv{}F_3^a\frac{\tau^3}{2\pi{}e}\lambda^3\int{}d\omega{}I_D(V_m-\hbar\omega/2e)
\omega [4+5 (\omega\tau)^2 + (\omega\tau)^4]^{-1}$ for the
symmetric part.% Dotted line shows the line $I=F_2^a I_N=F_3^a
%I_N$.
} \label{fig:antisymcur} \label{fig:symmcur}
\end{figure}

Let us now analyze $\delta I_A(V_m)$ for the noise source shown in
Fig.~\ref{fig:circuit}. The frequency dependence of the
nonsymmetrized noise $S_I(\omega)$ is derived, e.g., in
Ref.~\ocite{aguado00}. This derivation holds provided that the
Thouless energy $E_T$ for the resistors greatly exceeds $e\VS$.
Subtracting the fluctuations at $\VS=0$ we get the excess noise
$S_{2I}^i(\omega)=F_2^i {\cal S}(V_i)/R_i$, where $i=\{a,b\}$,
$V_i=R_i \VS/(R_a+R_b)$ and
\begin{equation}
\begin{split}
{\cal S}(V)&=\frac{eV\sinh(\frac{eV}{kT})-2\hbar \omega
\coth(\frac{\hbar \omega}{2kT}) \sinh^2(\frac{
eV}{2kT})}{\cosh(\frac{eV}{kT})-\cosh(\frac{\hbar \omega}{kT})}.
\end{split}
\label{eq:excessnoise}
\end{equation}
An example of the frequency dependence of $S_{2I}$ is plotted in
Fig.~\ref{fig:detectorband}. Note that it is symmetric with
respect to the sign reversal of $\omega$. This property can be
traced to the fact that in our example the source impedance,
characterizing quantum fluctuations, stays constant. Now, $S_{2I}$
can be substituted to Eq.~\eqref{eq:antisymmcur} to find the
effect of the driven noise on the detector current. Figure
\ref{fig:antisymcur} shows an example of $\delta I_A(V_m)$, i.e.,
the probed shot noise, for a few bias voltages $\VS$. For $\VS \gg
\hbar/\tau$, shot noise is essentially white over the detector
bandwidth (Fig.~\ref{fig:detectorband}) and thus the signal is
linear in $\VS$. In this case, $I_A(V_m)/\IS$
(Fig.~\ref{fig:antisymcur} inset) depends only on the factor
$D_A(V_m)$, which can be related to the calibration measurement,
and the Fano factors $F_2^{a,b}$, which can thus be measured from
this curve.

Next, consider the detection of the third cumulant of current
fluctuations. Its frequency dependence has been described in
various limits for different systems in Ref.~\cite{galaktionov03}.
At $T=0$, the third cumulant at zero frequency is of the form
$S_{3I}=F_3e^2I_N$, and its dispersion occurs on the scale
$\omega_c=\min(eV/\hbar,E_T/\hbar,I/e)$. For $R_s C_m\ll{}RC_J$
and a frequency independent $S_{3I}$ within the detection
bandwidth, the resulting $K_\phi(\omega)$ at $T=0$ is given by
\begin{equation}
  K_\phi(\omega)
  =
  2 \tau \lambda^3 (F_3^b - F_3^a) \frac{\IS}{e/\tau}
  \frac{\omega \tau}{4 + 5(\omega \tau)^2 + (\omega \tau)^4}
  \,.
  \label{eq:kphi}
\end{equation}
This is an antisymmetric function of $\omega$, due to the simple
form assumed for $S_{3I}$, and its frequency scale is given by $1/
\tau$. With the knowledge of the detector calibration current
$I_D(V_m)$, plugging $K_\phi(\omega)$ into Eq.~\eqref{eq:symcur}
allows us to calculate the response of the symmetric detector
current $I_S(V_m)$ to the third cumulant in principle for any type
of resistors $R_a$ and $R_b$. A few examples of $I_S(V_m)$ are
shown in Fig.~\ref{fig:symmcur}. As for the second cumulant,
following $I_S(V_m)/D_S(V_m)$ allows to measure the Fano factors.

\begin{figure}[h]
\centering
\includegraphics[width=\columnwidth]{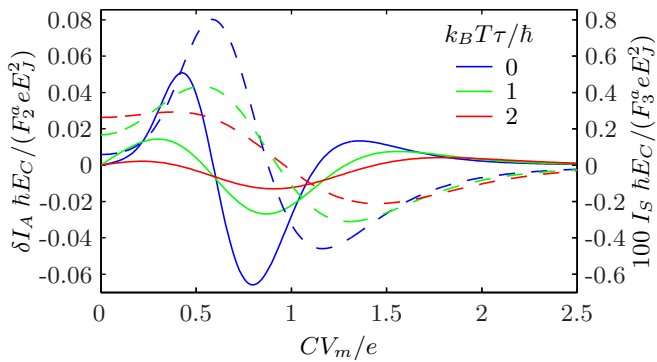}
\caption{(Color online) Antisymmetric (solid) and symmetric
(dashed) detection currents $\delta I_A(V_m)$ and $I_S(V_m)$ for
the same parameters as in Fig.~\ref{fig:antisymcur} but with a few
different temperatures at fixed $\VS=2e/C$, i.e., $\IS=60 e/\tau$.
For $I_S(V_m)$, we neglect the temperature dependence of the
intrinsic fluctuations.} \label{fig:tempdep}
\end{figure}

The general measurement scheme is valid at a finite temperature as
long as the Coulomb-blockade condition $T \lesssim E_C/k_B$ is
satisfied. The effects of $T>0$ are illustrated in
Fig.~\ref{fig:tempdep}. Compared to the previous example, two
corrections arise: the function $I_D(V_m)$ characterizing the
detector becomes smoother and its amplitude decreases, and the
form of the source excess fluctuations changes. The two are
characterized by different temperature scales, $E_C/k_B$ and
$e\VS/k_B$, respectively.

Measuring the calibration current $I_D(V_m)$ along with the
antisymmetric and symmetric currents, $\delta I_A(V_m)$ and
$I_S(V_m)$, allows to find the second and third cumulants of
excess current fluctuations in the source within the bandwidth
described in Fig.~\ref{fig:detectorband} (for typical parameters
\cite{sonin04}, in the range of 100 GHz) to an accuracy limited
mostly only by the resolution of the current measurement
(resolution of 0.1 pA yields $\sim$ 100 (fA)$^2$s for the second
and $\sim 0.01$ (fA)$^3$s$^2$ for the third cumulant
\cite{sonin04,deblock03}). In the limit $C_m \gg C_J$, $R_S C_m
\ll R C_J$, the only information required about the setup are the
resistances $R_a$, $R_b$ and $R$, and the sum capacitance $C_m +
C_J$, all of which can be measured separately. Thus, the scheme
allows an accurate determination of the second and third cumulants
over a large bandwidth and for the third cumulant, overcomes the
bandwidth problems encountered in Ref.~\cite{reulet03}.

In summary, we show in this Letter how incoherent Cooper-pair
tunneling in small JJs with a high-impedance environment can be
used for accurate and wide-band detection of voltage fluctuations.
Via the symmetry of the detector output current, one can identify
the contributions from the second and third cumulants separately.
While the presented example is on the measurement of fluctuations
in samples exhibiting no Coulomb blockade, this is not a
limitation of the scheme itself.

Note added in proof: During the refereeing process, we became
aware of Ref.~\cite{lesovik03}, which points out that the Fano
factors $F_3$ for the third cumulant depend on the definition of
the measured observable. Given that $F_3$ are related to the
nonsymmetrized observable, our results remain valid.

We acknowledge the useful discussions with W. Belzig, P. J.
Hakonen, G.-L. Ingold, M. Kindermann, R. Lindell and E. B. Sonin
and the support by EU-IHP ULTI III (HPRI-1999-CT-00050) visitor
program.


\begin{thebibliography}{99}
\bibitem{expansionnote} The validity of this
approximation can be traced back to the convergence of the
cumulant expansion and to the requirement $\tilde{S}_{2\phi}(0)
\ll 1$. The previous is valid provided that $\lambda \delta \omega
\tau < 1$, and the latter if $\IS \lambda^2 (F_2^a+F_2^b) \pi
\delta \omega \tau <  e/\tau$. Here $\delta \omega =
\min(\omega_b,1/\tau)$ and $2\pi\omega_b$ is the bandwidth of
$S_{2I}(\omega)$. As seen in Fig.~\ref{fig:antisymcur}, the
expansion is valid for a wide range of currents $I_B$.
For more details, see \cite{virtanenup}.
\bibitem{neglectedpart} If $S_\phi(-\omega) \neq S_\phi(\omega)$,
one gets an additional term $I_D[V_m-E/(2e)] S_\phi(E/\hbar)
\coth[\beta(eV_m-E/2)]$ in Eq.~\eqref{eq:antisymmcur}. Similarly,
one obtains an extra term $I_D[V_m-E/(2e)] K_\phi(E/\hbar)
\coth[\beta(eV_m-E/2)$ in Eq.~\eqref{eq:symcur} in the case
$K_\phi(-\omega) \neq -K_\phi(\omega)$.
\end{thebibliography}
\end{document}